\documentclass[twoside,fleqn]{article}
\usepackage{epsfig}
\usepackage{espcrc2}
\newcommand{\bc} {\begin{center}}
\newcommand{\ec} {\end{center}}
\newcommand{\bq} {\begin{equation}}
\newcommand{\eq} {\end{equation}}
\newcommand{\bqa}{\begin{eqnarray}}
\newcommand{\eqa}{\end{eqnarray}}

\newcommand{\nn}{\nonumber}
\newcommand{\al}{\alpha}
\newcommand{\be}{\beta}
\newcommand{\ga}{\gamma}
\newcommand{\de}{\delta}
\newcommand{\ep}{\epsilon}
\newcommand{\ph}{\phi}

\newcommand{\ka}{\kappa}
\title{
\vskip -100pt
\mbox{} \hfill BI-TP 99/32\\
\mbox{} \hfill September 1999\\
\vskip 65pt
Further Evidence for an unstable H-Dibaryon\,?}
\author{I.~Wetzorke with F.~Karsch, E.~Laermann
\thanks{The work has been supported by the TMR network ERBFMRX-CT-970122
and the DFG under grant Ka 1198/4-1.}
\\\vskip 6pt
Fakult\"at f\"ur Physik, Universit\"at Bielefeld, D-33615 Bielefeld, Germany}
\begin{document}
\begin{abstract}
We present preliminary results for the mass
of the 6q flavor-singlet state ($J^P=0^+$, $S=-2$)
called H-dibaryon, calculated in quenched QCD
on $16^3$x30 and $24^3$x30 lattices with improved
gauge and fermion actions (Symanzik improvement,
Clover action). For both lattice sizes we applied
the fuzzing technique to enhance the overlap with
the ground state. We observe a H-mass above the
$\Lambda\Lambda$-threshold for strong decay.
The difference in mass, $m_H - 2m_\Lambda$,
increases with increasing lattice size.
\end{abstract}
\maketitle
\section{INTRODUCTION}
In 1977 a bound six-quark state ({\it uuddss}), the
H-dibaryon, was predicted in a bag-model calculation by
Jaffe [1]. This state is the lowest SU(3) flavor
singlet state with spin zero, strangeness -2 and $J^P=0^+$.
These dibaryons may play an important role as Bose condensate
in nuclear matter prior to the quark-hadron phase transition of
QCD at high density.
\\[3mm]
In the last twenty years many attempts to verify the existence
and stability of this particle were undertaken by means of various
methods. Perturbative calculations included spin-dependent
\mbox{q-q} interactions arising from one gluon exchange (OGE) [1,2],
instanton induced interactions (III) [3,4] and Goldstone boson
exchange (GBE) [5]. The attractive q-q interaction is controlled
via the potential $V_s \sim (\lambda^a_1\lambda^a_2)(s_1 s_2)$,
where $\lambda^a_i$ denote SU(3)
color (OGE) or flavor (III/GBE) generators. The resulting
H-masses scattered in a range of a few hundred MeV around the
2231 MeV \linebreak $\Lambda\Lambda$-threshold for strong decay.
\\[3mm]
The lattice attempts to calculate $m_H$ a decade ago [6,7]
gave contradicting results, but a new calculation [8] showed
that the q-q attraction \linebreak decreases with increasing volume.
Hence it has been suggested that the H-dibaryon is unbound
\linebreak (approx.~110 MeV above the threshold)
in the \linebreak infinite volume extrapolation.
\section{DETAILS OF THE SIMULATION}
Our spectrum calculation was performed in quenched QCD with Wilson
fermions on lattices of the size $16^3 \times 30$ and $24^3 \times 30$.
We generated 57 independent gauge field configurations for the
smaller lattice and 15 configurations so far for the larger lattice,
separated by 100 sweeps of 4 overrelaxation and one heatbath step each.
For the gauge sector we applied the tree-level improved (1,2) Symanzik action
at a gauge coupling $\be=4.1$, which corresponds to a lattice spacing
$a=0.177(8){\rm\,fm}$ defined by the string tension [9].
%  \bqa
%  S_{(1,2)}\!\!\!&=&\!\!\!\be \sum_{x,\mu<\nu}~ {5\over 3}~\left(
%    1-\frac{1}{N}\re\tr\plaq_{\mu\nu}(x)\right)\nn \\[1mm]
%  &-&{1\over 6}\left(1-\frac{1}{2N}\re\tr
%    \left(\loOp_{\mu\nu}(x)+\lOop_{\mu\nu}(x)\right)\right)\nn
%  \eqa
A tree-level improved clover term was used in the fermionic part of the
action.
%  \bqa
%  S_C\!\!\!&=&\!\!\!\frac{1}{2 \kappa} \;\sum_{x,y}\; \bar{\Psi}(x) \left\{
%    \Bigg[\unit - \frac{\kappa}{2}\sum_{\mu,\nu}
%    \im\clover_{\!\!\mu\nu}\!(x)\,\sigma_{\mu\nu}(x) \Bigg]\delta_{x,y}
%  \right. \nn \\
%  & &\hspace*{-15mm}\left.-\;\kappa\;\sum_{\mu}\Bigg[
%    (\unit - \gamma_\mu )\,\delta_{x+\hat{\mu},y}\;\link_{\!\!\mu}(x) +
%    (\unit + \gamma_\mu )\,\delta_{x-\hat{\mu},y}\;\linkdag_{\!\!\mu}(y)
%    \Bigg] \right\} \Psi(y)
%  \nn
%  \eqa
\\[3mm]
The correlation functions were calculated via the relation
$C(t)=\langle{\cal O}(0){\cal O}^\dag(t)\rangle$
with appropriate operators for the lambda and the H-dibaryon [10]
(color indices: roman letters, spinor indices: greek letters):
\bqa
{\cal O}_\Lambda(x)\!\!\!&=&\!\!\!\ep_{abc} (C\ga_5)_{\be\ga}
[ u_{\al}^a(x) d_{\be}^b(x) s_{\ga}^c(x) \nn\\
&&\hspace*{-3.5mm}+d_{\al}^a(x) s_{\be}^b(x) u_{\ga}^c(x)
\!\!-\!2 s_{\al}^a(x) u_{\be}^b(x) d_{\ga}^c(x) ] \nn\\
{\cal O}_H(x)\!\!\!&=&\!\!\!3 (udsuds) -3 (ussudd) -3 (dssduu) \nn\\
({\sf abcdef})\!\!\!\!\!&=&\!\!\!\!\!\ep_{abc} \ep_{def}
(C\ga_5)_{\al\be} (C\ga_5)_{\ga\de} (C\ga_5)_{\ep\ph} \nn\\
&&\;*\;{\sf a}_{\al}^a(x) {\sf b}_{\be}^b(x) {\sf c}_{\ep}^c(x)
{\sf d}_{\ga}^d(x) {\sf e}_{\de}^e(x) {\sf f}_{\ph}^f(x) \nn
\eqa
We summed over spatial coordinates to project onto zero
momentum and over all spinor and color indices to get
diagonal correlators.
\\
Quark propagators were calculated at five different values
of the bare quark mass, which have been estimated to be in the range
50~-~250 MeV. The mass is controlled through the hopping parameter
$\kappa_u$ for the degenerated u- and d-quark mass and $\kappa_s$
for the s-quark mass. The physical $\kappa_u$ used in the
extrapolation was set by the ratio of nucleon mass and string
tension. The value for $\kappa_s$ was determined by the mass ratio
of lambda and nucleon.
\\
In addition to the H-dibaryon and lambda we calculated the
correlation functions for the strange mesons $K$, $K^*$ and the
$\Sigma$ baryon. The obtained particle masses lie all in a 10\%
range around the experimental values, which is a common result
in calculations using the quenched approximation.
\section{FUZZING}
Extended operators and smeared gauge fields provide
a better overlap with the ground state [11].
For the quark fields we use:
\bqa
\Psi'(x,R)=\!\!\!\!\sum_{\mu \in V_z}\!\!
(\,U^{\dagger}(x-\hat{\mu})...U^{\dagger}(x-R\hat{\mu})\Psi(x-R\hat{\mu})
\nn\\
\;\;\;\;+\;
U(x) ... U(x+(R-1)\hat{\mu})
\Psi(x+R\hat{\mu})) \nn
\eqa
Smeared gauge fields improve the simulation of the gluon cloud around the
quarks:
\bqa
U'(x,\mu)={\cal P}_{SU(3)}
\big[\;c\;U(x,\mu)+
\!\!\!\sum_{\pm\nu\not=\mu<4}
\!\!\!U_{staples(\nu)}\big] \nn
\eqa
\begin{figure}[t]
\bc
\epsfig{file=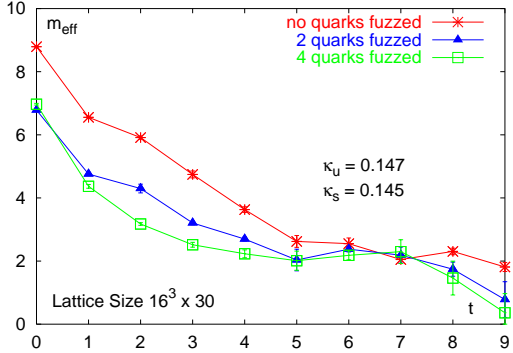,width=70.8mm}
\vskip -0.7truecm
\caption{Influence of fuzzing on the dibaryon correlation function
through different number of fuzzed quark propagators}
\label{fig:m_eff}
\vskip -0.4truecm
\ec
\end{figure}
The application of fuzzing for two of the six quarks inside the
dibaryon flattens the curvature of the effective mass (Fig.~1).
The largest plateau in the region with small errors is obtained
with fuzzed quarks for the four lighter u- and d-quarks.
Therefore we used this variant to calculate our correlation functions.
\section{RESULTS}
\subsection{Comparison of Dibaryon and Lambda Masses}
The particle masses were extracted from the long range behavior of the
correlation functions. We leave out successively data points in the
exponential fits at small t to yield stable results. \linebreak
Contributions of excited states were substantially reduced by
the fuzzing technique. Additionally, fits with two exponentials enlarge
the plateau even for the first time slices.
\begin{figure}[t]
\bc
\epsfig{file=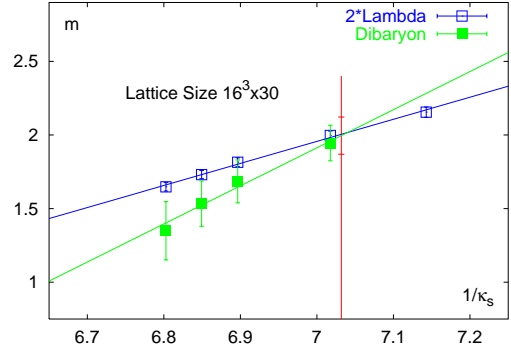,width=71mm}
\vskip -0.7truecm
\caption{Lambda and H-masses extrapolated to the
physical $\ka_u\!\!=\!0.149$ (line indicates physical $\ka_s$)}
\label{fig:cmp16}
\vskip -0.4truecm
\ec
\end{figure}
\\[3mm]
The results on the smaller lattice show a H-mass in the same range
as twice the $\Lambda$-mass (Fig.~2). We obtained $m_H$ = 2221(141)
MeV, which is just below the $\Lambda\Lambda$-threshold for strong decay.
For the difference in mass we obtain $m_H-2 m_\Lambda=$ \mbox{$-10\pm141$ MeV}.
It is therefore impossible to decide, whether the dibaryon is
unbound or slightly bound. Moreover, we cannot rule out
considerable finite size effects on the small
(2.8 fm)$^3$ \linebreak lattice.
\\[3mm]
The mass splitting on the larger lattice (4.2 fm)$^3$ seems to be
more pronounced (Fig.~3). The H-mass is larger than two
$\Lambda$'s for the parameter combinations calculated so far.
Although our statistics on the large lattice is not yet
sufficient to draw a definite conclusion, the preliminary results shown
in Fig.~3 suggest that the H-dibaryon is unbound in the infinite
volume limit.
\begin{figure}[t]
\bc
\epsfig{file=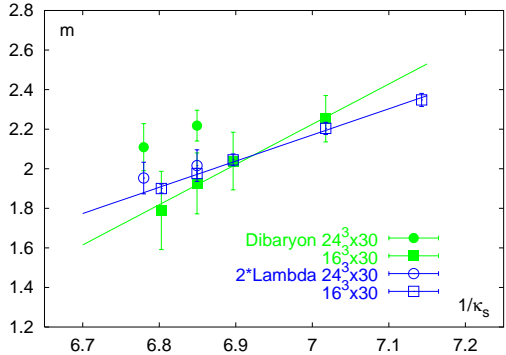,width=71mm}
\vskip -0.7truecm
\caption{Lambda and dibaryon masses
on both lattices for $\ka_u=0.1475$}
\label{fig:cmp24}
\vskip -0.4truecm
\ec
\end{figure}
\subsection{Ratio of the Correlation Functions}
The errors in the correlation functions for the dibaryon and the lambda
are strongly correlated. Some of the statistical uncertainties can thus
be eliminated by analysing directly the ratio of correlation functions.
The ratio of the correlation functions of dibaryon and lambda
should rise, when $m_H < 2 m_\Lambda$:
\bqa
C(H) \sim e^{-m_H t}
\;\;\Longrightarrow\;\;
\frac{C(H)}{C^2(\Lambda)} \sim e^{(2 m_\Lambda - m_H)t},\nn
\eqa
or should be constant, when $m_H > 2 m_\Lambda$:
\bqa
C(H) \sim e^{-2 m_\Lambda t}
\;\;\Longrightarrow\;\;
\frac{C(H)}{C^2(\Lambda)} \sim const.\nn
\eqa
On the smaller lattice this ratio is rising at $t\ge8$
with and without fuzzing (Fig.~4), which provides
evidence of a slightly bound H-dibaryon. On the larger lattice
the ratio remains constant until $t=8$.
With our present statistics the signal gets lost for larger values
of $t$. Nonetheless, the large plateau
of constant values again suggests that we see an
unbound H-dibaryon on this larger lattice.
\begin{figure}[t]
\bc
\epsfig{file=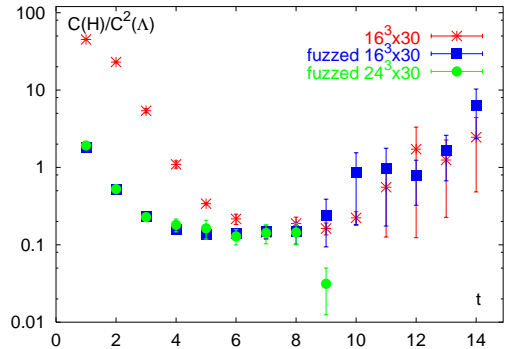,width=70.8mm}
\vskip -0.7truecm
\caption{Ratio of the correlation functions for
  $\ka_u=0.147$ ($0.1475$ for $24^3$x30) and $\ka_s=0.146$}
\label{fig:ratio}
\vskip -0.4truecm
\ec
\end{figure}
\section{CONCLUSIONS \& OUTLOOK}
With our present, still preliminary, analysis of the H-dibaryon
and $\Lambda$ masses we confirm the findings of Negele
et al.\,[8] presented at last year's lattice conference. We observe
that the difference in mass, $m_H-2m_\Lambda$, increases with
increasing lattice size and therefore will lead to an unbound
H-dibaryon in the infinite volume limit.
\\
We will continue our simulations on the larger lattice
with more configurations and additional combinations of
$\ka_u$ and $\ka_s$
to extrapolate to the physical values of the masses.
Simulations on a smaller lattice would be advantageous to confirm
the connection between finite lattice size and
the size of the mass splitting $m_H-2 m_\Lambda$.
\end{document}